\documentclass[3p]{elsarticle}

\journal{Journal of Computational Physics}

\usepackage{hyperref}
\usepackage{lineno}
\usepackage[title]{appendix}
\usepackage{amsmath}
\usepackage{dsfont}
\usepackage{array}
\usepackage{amssymb}
\usepackage{bm}
\usepackage{algorithm}
\usepackage{algpseudocode}
\usepackage{enumerate}
\usepackage{subcaption}
\usepackage{paralist}
\usepackage{multirow}
\usepackage{outlines}
\usepackage{todonotes}
\hypersetup{
	colorlinks,
	linkcolor={blue!50!black},
	citecolor={blue!50!black},
	urlcolor={blue!80!black},
	anchorcolor = {blue!80!black},
	filecolor = {blue!80!black},
	menucolor = {blue!80!black},
	runcolor = {blue!80!black}
}
\newcommand{\comment}[1]{}

\newcommand*{\ptder}[1]{\ensuremath{\frac{\partial {#1}}{\partial \tim}}} 
\newcommand{\tim}{\ensuremath{t} } 
\begin{document}

\begin{frontmatter}

\title{Diffuse interface treatment in generalized curvilinear coordinates with grid-adapting interface thickness}

\author[myaddress]{Henry Collis\corref{mycorrespondingauthor}}
\ead{hcollis@stanford.edu}
\cortext[mycorrespondingauthor]{Corresponding author}
\author[myaddress]{Shahab Mirjalili}
\ead{ssmirjal@stanford.edu}
\author[myaddress]{Ali Mani}
\ead{alimani@stanford.edu}
\address[myaddress]{Department of Mechanical Engineering, Stanford, CA 94305, USA}

\begin{abstract}
A general approach for transforming phase field equations into generalized curvilinear coordinates is proposed in this work. The proposed transformation can be applied to isotropic, non-isotropic, and curvilinear grids without adding any ambiguity in determining the phase field parameters. Moreover, it accurately adapts the interface thickness to the local grid-size for a general curvilinear grid without creating oscillations. Three canonical verification tests are presented on four grids with varying skewness levels. The classic advection and drop in shear tests are extended to curvilinear grids and show that the original phase field on Cartesian grids and the proposed curvilinear form have an identical order of convergence. Additionally, the proposed method is shown to provide grid-independent convergence on a two-way coupled compressible Rayleigh-Taylor instability. These simulations illustrate the robustness and accuracy of the proposed method for handling complex interfacial structures on generalized curvilinear grids. 
\end{abstract}


\end{frontmatter}

\section{Introduction} \label{sec:introduction}

Two-phase flows are common in nature and have many applications in engineering. Phase field methods have emerged as a popular approach for solving two-phase flow problems \citep{Mirjalili_ARB,Roccon2023}. There are multiple families of phase field models including those based on the Cahn-Hilliard \cite{cahn1958} and Allen-Cahn equations \cite{allen1979}, in addition to more recent models based on the conservative form of the Allen-Cahn equation, first proposed by \cite{Chiu2011}. This latter category of models is also referred to as the conservative second-order phase field and conservative diffuse interface (CDI) method. With recent advancements, consistent models based on all of these phase field models can robustly handle complex interfacial flows with large density ratios \citep{Huang2020,khanwale2020, Mirjalili2021}. In this work, we will extensively discuss the CDI model, which has gained popularity in recent years. There has been significant progress in recent years on the CDI method, including discretization schemes that provide bounded volume fractions \cite{Mirjalili2020}. Consistently applying CDI model to the energy and momentum equations has also been shown to enable robust and accurate simulations of high-density ratios at high Re numbers for incompressible and compressible two-phase flows \cite{Mirjalili2020,Jain2020}. Recently, the CDI method has been extended to handle N-phase immiscible flows while retaining the attractive properties described above \cite{mirjalili2024}. Additionally, the CDI form combined with high-order shock-capturing schemes has been shown to be highly effective in modeling compressible two-phase flows involving shocks and other multiphase compressibility effects \cite{Collis2022,Jain2023}. 

The phase field methods discussed above have been extensively studied on isotropic Cartesian grids. While isotropic Cartesian grids are useful for many settings, they cannot be readily used for simulations of flows in complex geometries. Unstructured grids have proven to be a general and effective approach for general geometries. Extensions for the Cahn-Hilliard, Allen-Cahn, and recently the CDI phase field methods have been tested on both unstructured grids and adaptive octree grids using finite element and finite volume discretizations \cite{chiapolino2017sharpening,joshi2018_CAC_unstructured,khanwale2020,khanwale2023,hwang2024robust}. Unstructured grids provide a flexible way to handle complex geometries but they come with many challenges, including higher computational costs and increased complexity compared to a structured counterpart. A generalized curvilinear formulation provides an avenue to use standard discretization schemes to handle complex geometries. Moreover, when combined in a multi-block form they can capture advanced geometries while retaining the advantages of structured grids such as cost-efficiency, algorithmic simplicity, and amenability to high-order spatial discretizations \cite{steinthorsson1993development,chima2003swift}.

The objective of this work is to provide stable and accurate curvilinear formulations for phase field methods. By adopting the CDI formulation as a use case, we show how the direct transformation of phase field methods into curvilinear coordinates causes ambiguity in parameter selection which can lead to interface oscillations or overly thick interfaces. To avoid these issues, we provide a systematic approach for deriving the curvilinear extension of phase field methods, resulting in stable and accurate phase interfaces on curvilinear grids. Specifically, the proposed methodology results in interfaces that automatically adapt to the local grid such that the number of grid points across the interface is approximately the same everywhere in the domain.

The rest of the paper is organized as follows. Section \ref{sec:PhaseFieldEq} introduces the adopted CDI phase field approach and illustrates the advantages of the proposed curvilinear transformation compared to other approaches. Section \ref{sec:Results} provides three verification tests to show the convergence and stability of the proposed curvilinear transformation. These include a curvilinear droplet advection, a drop in shear flow on curvilinear grids, and a Rayleigh-Taylor instability. Finally, the conclusions of this work are provided in Section \ref{sec:Conclusion}.

\section{Phase field equation} \label{sec:PhaseFieldEq}
\begin{figure}[hbt!]
\begin{center}
\includegraphics[width=0.7\textwidth]{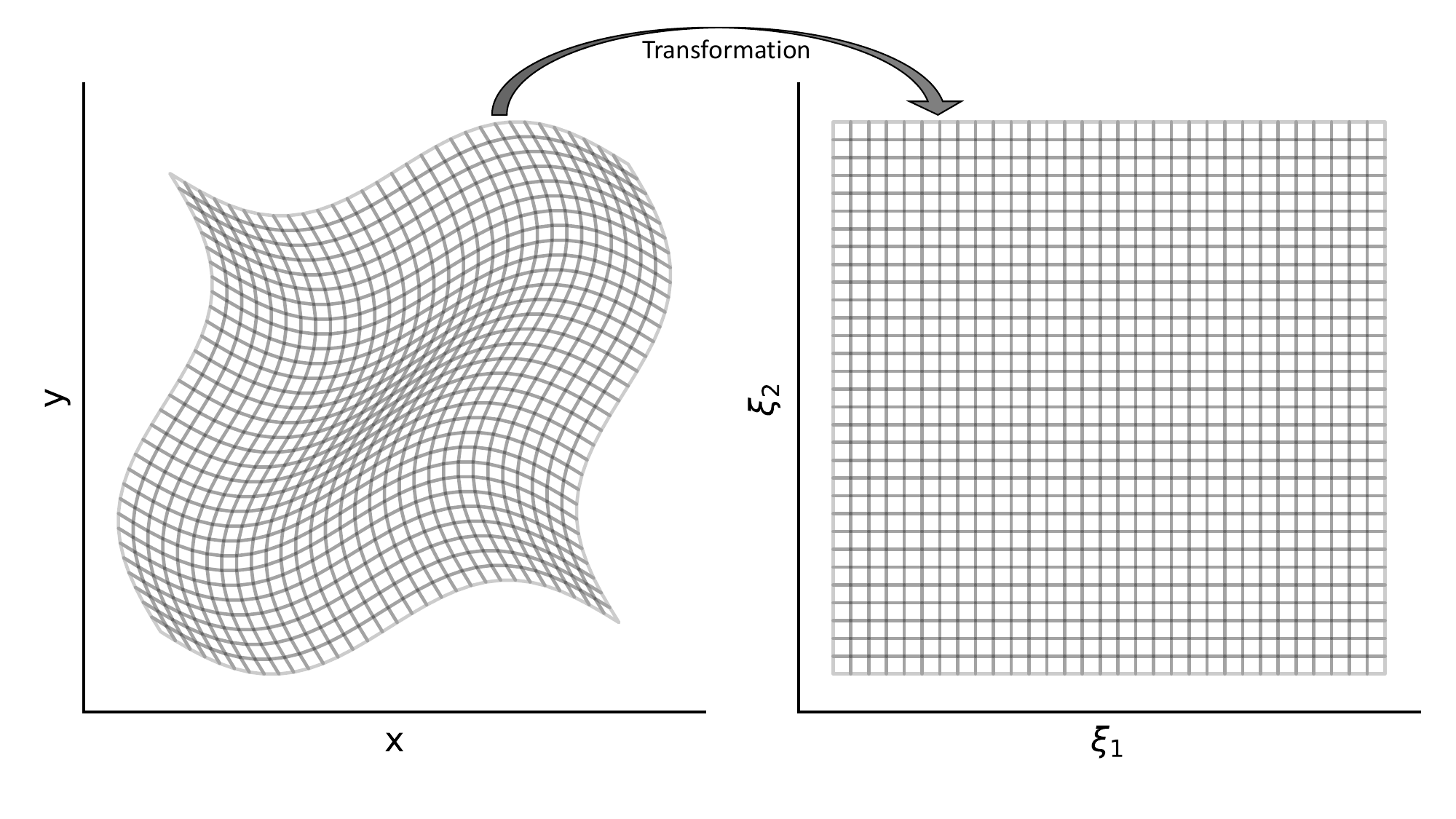}
\caption{An example of a curvilinear domain in physical space is shown in the left panel and the corresponding computational domain in the right panel. The transformation between these grids was completed using the Grid 4 projection in Table \ref{tab:grid_transformation}. The grid metrics project the system from physical space $(x,y)$ to computational space $(\xi_1,\xi_2)$ so it can be solved on a isotropic computational grid. \label{fig:Cart_to_Curv}}
\end{center}
\end{figure}

A general form for a phase field equation in cartesian coordinates is
\begin{equation} 
    \ptder{\phi} + \frac{\partial (u_i \phi)}{\partial x_i} = g(\phi,x_i), 
    \label{eq:PF_cartesian}
\end{equation} 
where $\phi$ is the phase field variable (volume fraction) ranging from $[0,1]$,with $\phi=0$ and $\phi=1$ representing the pure phases (phase $1$ and $2$, respectively), $u_i$ is the advection velocity in physical space, and $x_i$ are the principle axis for the grid in physical space. In this form, $g(\phi, x_i)$ represents a general phase field approach and is inclusive of Cahn-Hilliard, Allen-Cahn, and CDI-type models. Directly transforming this equation to generalized curvilinear coordinates results in
\begin{equation} 
    \frac{1}{J}\ptder{\phi} + \frac{\partial (U_i \phi/J)}{\partial \xi_i} = G(\phi, \xi_i),
    \label{eq:PF_curvilinear}
\end{equation}
where the contravariant velocity components are defined as, $U_i = u_j({\partial \xi_i}/{\partial x_j})$, and $J$ is the determinant of the metric Jacobian, where ${1}/{J} = \det\left( {\partial x_i}/{\partial \xi_j}\right)$. The grid metrics (${\partial \xi_i}/{\partial x_j}$ and $J$) define a mapping between a physical grid ($x,y$) and a computational ($\xi_1,\xi_2$) grid as shown in Figure \ref{fig:Cart_to_Curv}. 

This work proposes a strategy for determining $G(\phi, \xi_i)$ for phase field methods. To illustrate the proposed curvilinear  transformation for the CDI method is carried out in detail here. The proposed curvilinear form for other phase field methods are listed in \ref{appx:transformations}. On uniform grids CDI takes the form in Eq. \ref{eq:CDI_cartesian}.

\begin{equation} \label{eq:CDI_cartesian}
    g(\phi, x_i) = \frac{\partial}{\partial x_i}\left[\Gamma\left(\epsilon\frac{\partial \phi}{\partial x_i} - \phi(1-\phi)\frac{\frac{\partial \phi}{\partial x_i}}{(\frac{\partial \phi}{\partial x_k}\frac{\partial \phi}{\partial x_k})^{1/2}}\right)\right]
\end{equation}
Two parameters, $\Gamma$ and $\epsilon$, are velocity and length scales respectively. For uniform grids, previous works have provided criteria for selecting these model parameters to guarantee the volume fraction, $\phi$, remains stable and bounded \cite{Mirjalili2020, Jain2020}. Following these works, $\Gamma$ is chosen to be proportional to the max velocity in the domain, and the length scale, $\epsilon$, is proportional to the isotropic grid spacing, $\Delta$. 

One of the advantages of the equation based interface regularization models (like CDI) is the explicit PDE form enables analytical analysis as well as wide applicability for many discretization schemes. Along these lines, it is reasonable to expect that the PDE form of CDI can be transformed into a generalized curvilinear coordinate system and perform well as a model. This transformation will be known as the ``direct'' transformation and the issues associated with this approach will be explained in the following section. 

\subsection{Direct Curvilinear Transformation} \label{sec:DirectCurvilinear}
\begin{figure}[hbt!]
\begin{center}
\includegraphics[width=\textwidth]{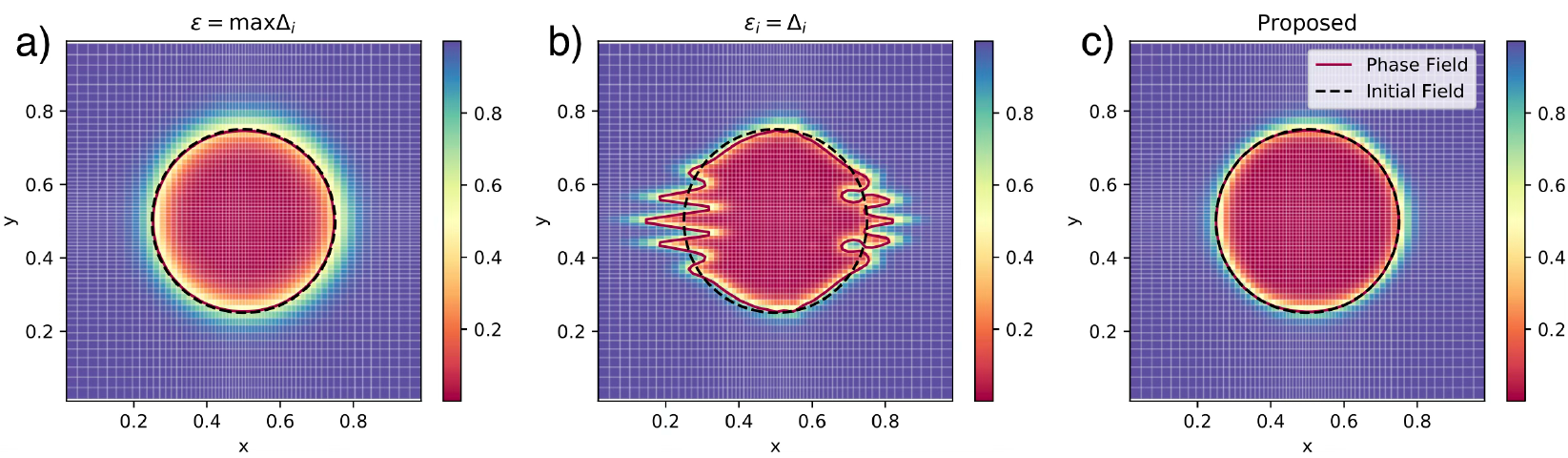}
\caption{The final shapes of a circular drop after 1 period of advection through a non-uniform rectilinear grid for two curvilinear projection methods. Panel (a) shows the results for the direct projection to curvilinear coordinates using option 1: $\epsilon = \max_i\Delta_i$. Panel (b) shows the results for the direct projection to curvilinear coordinates using option 2: $\epsilon_i = \Delta_i$. Panel (c) shows the results from the proposed curvilinear projection. \label{fig:Comp_CurvForms}}
\end{center}
\end{figure}

The first strategy for transforming the phase field regularization terms is directly transforming the PDE form into generalized curvilinear coordinates. This strategy will be called the ``direct" transformation in this work and is detailed in \ref{appx:OperatorCurvilinearTransformation}.  Applying the direct transformation to the CDI terms results in a form for $G(\phi, \xi_i)$ given by

\begin{equation} \label{eq:CDI_curvilinear}
    G(\phi, \xi_i) = \frac{\partial }{\partial \xi_i}\left[\Gamma\left(\epsilon \frac{\partial \phi}{\partial \xi_k}\frac{\partial \xi_k}{\partial x_j} - \phi(1-\phi)\frac{\frac{\partial \phi}{\partial \xi_k}\frac{\partial \xi_k}{\partial x_j}}{\left(\frac{\partial \phi}{\partial \xi_k}\frac{\partial \xi_k}{\partial x_q}\frac{\partial \phi}{\partial \xi_l}\frac{\partial \xi_l}{\partial x_q}\right)^{1/2}}\right)\frac{\partial \xi_i}{\partial x_j}/J\right].
\end{equation}
In the direct transformation, the velocity scale, $\Gamma$, will remain proportional to the maximum physical velocity in the domain, $\Gamma = \max_i|u_i|$. However, because the interface thickness needs to be resolved by the grid, determining the interface thickness parameter, $\epsilon$, is not straight-forward as the grid-size is no longer uniform. To illustrate this, for a rectilinear non-uniform grid two options for defining $\epsilon$ include, 
\begin{enumerate}
  \item uniform interface thickness: $\epsilon = \max_i \Delta_i$,
  \item non-uniform interface thickness: $\epsilon_i = \Delta_i$.
\end{enumerate}
Option 1 results in a constant thickness phase interface in physical space, but since the interface thickness needs to be resolved by the grid, it has to be on the order of the largest grid size in the domain. Thus, this option does not take advantage of the local grid refinement allowed by both non-uniform rectilinear grids and more general curvilinear grids. Option 2 redefines $\epsilon$ as a vector, $\epsilon_i$, dependent on the local grid size in the direction $i$ and has been used as a strategy in a recent work applying the conservative phase field regularization terms to unstructured grids \cite{hwang2024robust}. Both choices for $\epsilon$ can be tested on a non-uniform rectilinear grid defined using grid stretching. The stretched grid is defined by mapping a uniform computational domain with $\xi_1=[-1,1]$ and $\xi_2=[-1,1]$ to a non-uniform grid using Eq. \ref{eq:grid_stretching} with the stretching factor, $S = 0.2$. 

\begin{equation} \label{eq:grid_stretching}
        x = \frac{1}{2}\left(\frac{\sinh(S\xi_1)}{\sinh(S)} + 1\right),  \quad
        y = \frac{1}{2}\left(\frac{\sinh(S\xi_2)}{\sinh(S)} + 1\right) 
\end{equation}
A drop is initialized using Eq. \ref{eq:IC_advect_physical} with $x^c = y^c = 0.5$, $N_x = N_y = N=100$, $R = 0.25$ and is advected with $\Vec{u}=(1,0)$ in a periodic $1\times1$ domain for one period using the numerical method described in Section \ref{sec:NumericalMethods}.

\begin{equation} \label{eq:IC_advect_physical}
    \phi(x,y) = \frac{1}{2}\left(1-\tanh{\left(R - \sqrt{\frac{(x-x^c)^2+(y-y^c)^2}{2\times(2/N)}}\right)}\right)
\end{equation} 
The results of the simulations for the direct application of CDI using options 1 and 2 on the non-uniform rectilinear grid described in Eq. \ref{eq:grid_stretching} are shown in Figure \ref{fig:Comp_CurvForms} (a) and (b), respectively, both revealing undesirable characteristics. Panel (a) in Figure \ref{fig:Comp_CurvForms} illustrates how the uniform interface thickness given by option 1 is much larger than required by the local grid. Panel (b) in Figure \ref{fig:Comp_CurvForms} shows that option 2 results in a highly distorted interface shape caused by the directional-dependent non-uniform $\epsilon_i$ and its interaction with the nonlinear sharpening term. Panel (c) in Figure \ref{fig:Comp_CurvForms} shows a desirable phase interface which locally refines with the grid without distorting the drop shape. The phase field result in Panel (c) of Fig. \ref{fig:Comp_CurvForms} is obtained using our proposed curvilinear transformation applied to the CDI model. This transformation is presented in the following section.

\subsection{Proposed Curvilinear Transformation} \label{sec:Proposed_curvilinear_form}

As shown in Section \ref{sec:DirectCurvilinear}, transforming the operators in Eq. \ref{eq:CDI_cartesian} from Cartesian to curvilinear form resulted in ambiguity for determining the interface thickness parameter $\epsilon$ for non-isotropic grids. Instead, the phase field model can be defined directly in computational space where the grid is isotropic. For example, the CDI model given by Eq. \ref{eq:CDI_cartesian} can be defined directly in computational space by substituting the physical coordinate system with the computational coordinate system, 
\begin{equation} \label{eq:CDI_computational}
\Gamma\left(\epsilon\frac{\partial \phi}{\partial x_i} - \phi(1-\phi)\frac{\frac{\partial \phi}{\partial x_i}}{(\frac{\partial \phi}{\partial x_k}\frac{\partial \phi}{\partial x_k})^{1/2}}\right) \rightarrow \Gamma_{\xi}\left(\epsilon_{\xi}\frac{\partial \phi}{\partial \xi_i} - \phi(1-\phi)\frac{\frac{\partial \phi}{\partial \xi_i}}{\left(\frac{\partial \phi}{\partial \xi_k}\frac{\partial \phi}{\partial \xi_k}\right)^{1/2}}\right).
\end{equation}


One advantage of defining the CDI terms directly in the computational space is that the computational domain is defined with uniform grid spacing $\Delta\xi_i = \Delta_{\xi}$. As such, similar to Eq. \ref{eq:CDI_cartesian}, $\epsilon_{\xi}$ can be defined to be proportional to the computational grid spacing $\Delta_{\xi}$. The balance of diffusion and sharpening fluxes in the computational grid results in equilibrium interface profiles with constant thickness $(\Delta_{\xi})$ in the computational space. Additionally, the velocity parameter, $\Gamma_{\xi}$, can be defined as the maximum contravariant velocity, $\Gamma_{\xi} = \max_i|U_i|$. 

To apply the transformation in Eq. \ref{eq:CDI_computational} to an interface in physical space, the computational CDI flux can be projected back to physical space using 
\begin{equation} \label{eq:CDI_comp_physical}
    \Theta_j = \Gamma_{\xi}\left(\epsilon_{\xi}\frac{\partial \phi}{\partial \xi_i} - \phi(1-\phi)\frac{\frac{\partial \phi}{\partial \xi_i}}{\left(\frac{\partial \phi}{\partial \xi_k}\frac{\partial \phi}{\partial \xi_k}\right)^{1/2}}\right)\frac{\partial x_j}{\partial \xi_i}. 
\end{equation}
At this point, $\Theta_j$ is a vector in physical space and the outer-divergence of Eq. \ref{eq:CDI_cartesian} can be projected back to computational space using the rules described in \ref{appx:OperatorCurvilinearTransformation} to obtain,
\begin{equation} \label{eq:CDI_comp_curv_notsimplified}
    G(\phi,\xi_i) = \frac{\partial }{\partial \xi_i}\left[\Theta_j\frac{\partial \xi_i}{\partial x_j}/J\right] = \frac{\partial }{\partial \xi_i}\left[\Gamma_{\xi}\left(\epsilon_{\xi}\frac{\partial \phi}{\partial \xi_k} - \phi(1-\phi)\frac{\frac{\partial \phi}{\partial \xi_k}}{\left(\frac{\partial \phi}{\partial \xi_q}\frac{\partial \phi}{\partial \xi_q}\right)^{1/2}}\right)\frac{\partial x_j}{\partial \xi_k}\frac{\partial \xi_i}{\partial x_j}/J\right],
\end{equation}
where we can observe that $\frac{\partial x_j}{\partial \xi_k}\frac{\partial \xi _i}{\partial x_j} \equiv I$ to obtain the final form of the proposed CDI model in generalized curvilinear coordinates in Eq. \ref{eq:CDI_comp_curv}.
\begin{equation} \label{eq:CDI_comp_curv}
    G(\phi,\xi_i) = \frac{\partial }{\partial \xi_i}\left[\Gamma_{\xi}\left(\epsilon_{\xi}\frac{\partial \phi}{\partial \xi_i} - \phi(1-\phi)\frac{\frac{\partial \phi}{\partial \xi_i}}{\left(\frac{\partial \phi}{\partial \xi_j}\frac{\partial \phi}{\partial \xi_j}\right)^{1/2}}\right)/J\right]
\end{equation}

The form of $G(\phi,\xi_i)$ in Eq. \ref{eq:CDI_comp_curv} can be applied to both curvilinear and rectilinear grids without ambiguity in parameter selection. For an isotropic Cartesian grid, the proposed model is identical to the original model form in Eq. \ref{eq:CDI_cartesian}. The application to the non-uniform rectilinear grid defined by Eq. \ref{eq:grid_stretching} is shown Panel (c) of Fig. \ref{fig:Comp_CurvForms}. In general, the proposed model naturally adapts the interface thickness to the physical grid while preserving an accurate interface shape. The regularity of the interface on the computational grid is transformed using the grid metrics to achieve a well-behaved, adapting interface in physical space. Additionally, the boundedness properties illustrated in earlier works on uniform, isotropic Cartesian grids will also hold for generalized curvilinear grids with appropriate choices for parameters $\epsilon_{\xi}$ and $\Gamma_{\xi}$ as illustrated in Section \ref{sec:boundedness}.

\subsubsection{Extensions to other regularization forms} \label{sec:PhaseField_extensions}
A similar derivation can be applied to other forms of $g(\phi, x_i)$. For instance, the phase field variable $\phi$ can be transformed into an approximate signed distance function $\psi$ using the relationship,
\begin{equation} \label{eq:psi_transformation}
    \psi = \epsilon \ln{\frac{\phi + \delta}{1-\phi +\delta}}
\end{equation}
where $\delta$ is a small number taken as $10^{-100}$ \cite{chiodi2017reformulation}. The signed distance function is a smooth field (approximately linear) through the interface, whereas $\phi$ tends to converge towards an interface with a $\tanh$ profile. Consequently, utilizing the $\psi$ field reduces numerical errors during the interpolations required for calculating the interface normal and the nonlinear term on cell faces \cite{chiodi2017reformulation}. The reformulation of the CDI model in terms of the signed distance function $\psi$ is known as the ACDI model \cite{Jain2022ACDI}. The corresponding ACDI formulation in generalized curvilinear coordinates is included in \ref{appx:transformations}. Both ACDI and CDI models are analytically equivalent at the PDE-level. However, they are distinguished by the beneficial numerical properties afforded by using interpolations based on the $\psi$ field. Given its enhanced numerical accuracy, the ACDI model will be used for the remaining test cases in this study.

\section{Numerical Methods} \label{sec:NumericalMethods}

To obtain the numerical results presented in this work, instead of only solving Eq. \ref{eq:PF_curvilinear} directly, we solve the four equation model for two phase flows which assumes thermal and mechanical equilibrium between phases \cite{kataoka1986local,shyue1998efficient,cook2009enthalpy}. Additionally, we augment the system by the regularization terms defined in the RHS of Eq. \ref{eq:PF_curvilinear}. The four equation model we solve consists of transport equations for the two phases, the momentum equation, and the total energy equation, with the addition of regularization terms \cite{Collis2022,Jain2023}. The system of equations in curvilinear coordinates coupled with the CDI regularization terms can be written in a compact form as,
\begin{equation*}
    \frac{1}{J}\frac{\partial \bm C}{\partial t} + 
      \frac{\partial (\bm F/J)}{\partial \xi_1} +
             \frac{\partial (\bm G/J)}{\partial \xi_2} +
             \frac{\partial (\bm H/J)}{\partial \xi_3}
     =  \frac{\partial (\bm F_{\nu}/J)}{\partial \xi_1} + \frac{\partial (\bm H_{\nu}/J)}{\partial \xi_3} +
             \frac{\partial (\bm G_{\nu}/J)}{\partial \xi_2} + \frac{\partial(\bm F_{DI}/J)}{\partial \xi_1} + \frac{\partial(\bm G_{DI}/J)}{\partial \xi_2} +
              \frac{\partial (\bm H_{DI}/J)}{\partial \xi_3}
\end{equation*}
where $\bm C$ is the vector of conserved variables, $\bm F$, $\bm G$ and $\bm H$ are the inviscid fluxes, $\bm F_{\nu}$, $\bm G_{\nu}$ and $\bm H_{\nu}$ are the viscous fluxes, and $\bm F_{DI}$, $\bm G_{DI}$ and $\bm H_{DI}$ are the phase field fluxes in the 3 directions $(\xi_1,\xi_2,\xi_3)$ respectively. As an example, the vector of the conserved variables, and the fluxes in the $\xi_1$ direction are provided below,
\begin{equation}
    \bm C =
    \begin{bmatrix}
    \rho Y_1 \\ \rho Y_2 \\
    \rho u \\ \rho v \\ \rho w \\ E
    \end{bmatrix}, \quad
    \bm F =
    \begin{bmatrix}
    \rho Y_1U_1  \\ \rho Y_2U_1 \\
    \rho U_1u + p \partial \xi_1/\partial x \\ \rho U_1v + p\partial \xi_1/\partial y \\ \rho U_1w + p\partial \xi_1/\partial z \\ U_1(E+p)
    \end{bmatrix}, \quad
    \bm F_{\nu}=
    \begin{bmatrix}
    0 \\ 0 \\
    \tau_{1i}\partial \xi_1/\partial x_i \\
    \tau_{2i}\partial \xi_1/\partial x_i\\
    \tau_{3i}\partial \xi_1/\partial x_i\\
    b_i\partial \xi_1/\partial x_i
    \end{bmatrix}, \quad
    \bm F_{DI} = \begin{bmatrix} 
    A_1  \\ A_2 \\ 
    u \Sigma_l A_l \\ 
    v \Sigma_l A_l \\ 
    w \Sigma_l A_l  \\ 
    k\Sigma_l A_l + \Sigma_l h_lA_{l} 
    \end{bmatrix}
\end{equation}
where $\rho$ is the mixture density, $Y_l$ is the mixture mass fraction, $u, v, w$ are the velocity components in physical space, $U_k$ are the contravariant velocity components, $\tau_{ij} = \mu\left[\frac{\partial u_i}{\partial x_j} + \frac{\partial u_j}{\partial x_i} - 2/3\frac{\partial u_k}{\partial x_k}\delta_{ij}\right]$, with $\mu$ as the dynamic viscosity, and $\delta_{ij}$ as the Kronecker delta. Additionally, $b_i=u_j\tau_{ij}$, $E = \rho(e + k)$ where $e$ is the mixture internal energy and $k$ is the kinetic energy, and $U_i = u_j\frac{\partial \xi_i}{\partial x_j}$. For this system heat conduction between phases is not included. $A_l$ is the phase field flux in the $\xi_1$ direction defined by scaling Eq. \ref{eq:CDI_computational} by phasic density and transforming the CDI model to the ACDI model using Eq. \ref{eq:psi_transformation} (for reasons described in Section \ref{sec:PhaseField_extensions}),
\begin{equation}
    A_l = \rho_l\left[\Gamma_{\xi}\left(\epsilon_{\xi}\frac{\partial \phi}{\partial \xi_i} - \frac{1}{4}\left[1-\tanh^2{\left(\frac{\psi_{\xi}}{2\epsilon_{\xi}}\right)}\right]\frac{\partial \psi_{\xi}/\partial \xi_i}{|\Vec{\nabla}\psi_{\xi}|}\right)/J\right].
\end{equation}
The full system of equations must be solved for the Rayleigh-Taylor problem described in Section \ref{sec:Results_RTI}, but for all other tests in Section \ref{sec:Results} the system is one-way coupled to the regularization flux terms and reduces to equivalently solving Eq. \ref{eq:PF_curvilinear}. The simplification of the entire system is achieved by defining both phases as identical fluids. Since the phases are identical, we can note that $\rho_1 = \rho_2$ and $h_1 = h_2$. This leads to $\sum_l A_l = 0$ which will remove all effects of regularization in the momentum and energy equations. Additionally, for the one-way coupled system, the pressure and temperature can be set as constant throughout space and time. Constant pressure and temperature lead to the density remaining constant and the equivalence between mass-fraction and volume fraction ($Y_l = \phi_l$). As such, the mass equation simplifies exactly to the volume fraction equation shown in Eq. \ref{eq:PF_curvilinear}. The final term needed to close the system is a prescription of the velocity field as a function of time which is specific to each test-case shown in Section \ref{sec:Results}. 

The convective terms are discretized using a 6th order skew-symmetric kinetic-energy and entropy-preserving (KEEP) scheme extended for multi-component flows on generalized curvilinear grids \cite{Kuya2021}. The viscous and phase field terms are discretized using 2nd order centered difference schemes. The phase field terms are implemented in a consistent manner as the convective terms to allow the system to retain the desired KEEP properties \cite{Jain2022KEEP}. A 3rd order strong stability preserving Runge-Kutta scheme was used for time integration \cite{gottlieb2001strong}. The above system is implemented into the highly parallel Hypersonic Task-based Regent (HTR) solver, which uses the Legion language for parallel task management on general heterogeneous computational architectures \cite{bauer2012legion,DiRenzo2020,di2021htr,di2022htr}. We have utilized this code and numerical scheme in combination with an ENO-type reconstruction on Cartesian grids in our previous work \citep{Collis2022}.

\section{Results} 
\label{sec:Results}

To examine the proposed curvilinear transformation, three numerical experiments are conducted. First, a drop advection problem is carried out on four grids, demonstrating that the proposed phase field equation on curvilinear grids provides accurate and bounded volume fraction fields. Next, a drop in shear flow is simulated to analyze the impact of curvilinear grids on shape convergence. Finally, we coupled the proposed curvilinear transformation for the phase field equations with the compressible Navier Stokes equations to simulate two-dimensional Rayleigh-Taylor instability on different grids. This experiment demonstrates how the method generalizes to a two-way coupled fluid dynamics problem and converges to identical physical solutions across diverse grid configurations. Three curvilinear mappings from computational to physical space are shown in Table \ref{tab:grid_transformation}, and the relevant parameters for each computational experiment are outlined in Table \ref{tab:grid_params}. 

\begin{table}[hbt!]
    \centering
    \begin{tabular}{l c c}
    \hline
    \\
     & Computational Domain & Physical Domain\\
    \\
    \hline
    \\
    Grid 1 &  $\begin{aligned}
        \xi_1 = [0,1], \quad 
        \xi_2 = [0,1]
    \end{aligned}$ & 
    $\begin{aligned}
        x &= w\xi_1, \quad 
        y = w\xi_2 
    \end{aligned}$ \\\\
    \hline
    \\
    Grid 2 &  $\begin{aligned}
        \xi_1 &= [0,1], \quad
        \xi_2 = [0,1]
    \end{aligned}$ & 
    $\begin{aligned}
        x &= w\left[\xi_1 + S\sin(\lambda \pi \xi_2)\right] \\
        y &= w\left[\xi_2 + S\sin(\lambda \pi \xi_1)\right]
    \end{aligned}$ \\\\
    \hline
    \\
    Grid 3 & $\begin{aligned}
        \xi_1 = [0,1], \quad
        \xi_2 &= [0,1]
    \end{aligned}$ &
    $\begin{aligned} 
        x &= w\left[\xi_1 + S\sin(\lambda \pi \xi_1)\sin(\lambda \pi \xi_2)\right] \\
        y &= w\left[\xi_2 + S\sin(\lambda \pi \xi_1)\sin(\lambda \pi \xi_1)\right]
    \end{aligned}$  \\
    \\
    \hline
    \\
    Grid 4 & $\begin{aligned}
        \xi_1 &= [0,1], \quad
        \xi_2 = [0,1]
    \end{aligned}$ & $\begin{aligned} 
        x &= w\left[\xi_1 + S\sin(\lambda \pi (2\xi_2-1))\right] \\
        y &= w\left[\xi_2 + S\sin(\lambda \pi (2\xi_1-1))\right]
    \end{aligned}$  \\
    \\
    \hline
    \\
    \end{tabular}
    \caption{Computational and physical domain mappings for three grid options. The parameter $S$ controls the skewness of the grid in physical domain and $\lambda$ controls the frequency of the sinusoidal oscillations.}
    \label{tab:grid_transformation}
\end{table} 

\begin{table}[hbt!]
    \centering
    \begin{tabular}{l c c c}
    \hline
    \\
     & Drop Advection  & Drop in Shear & RT-Instability\\
    \\
    \hline
    \\
    Grid 1 & 
    $w=1$& 
    $w=1$& 
    $w=3$  \\\\
    \hline
    \\
    Grid 2 & 
    $S = 1/80$, $\lambda = 8$, $w=1$& 
    Not applicable & 
    $S = 3/80$, $\lambda = 8/3$, $w=3$  \\\\
    \hline
    \\
    Grid 3 & 
    $S = 1/20$, $\lambda = 4$, $w=1$& 
    Not applicable& 
    $S = 1/10$, $\lambda = 2/3$, $w=3$  \\
    \\
    \hline
    \\
    Grid 4 & 
    $S = 1/10$, $\lambda = 1$, $w=1$&
    $\begin{aligned}
        S = 1/20 \text{, } \lambda = 1\text{, } w=1 \\
        S = 1/10\text{, } \lambda = 1\text{, } w=1
    \end{aligned}$& 
    $S = 1/10$, $\lambda = 1$, $w=3$  \\
    \\
    \hline
    \\
    \end{tabular}
    \caption{Required parameters for defining the curvilinear grids used for the computational experiments}
    \label{tab:grid_params}
\end{table}

\subsection{Drop Advection on Curvilinear Grids} \label{sec:Drop_Advection}

In this section, the proposed curvilinear projection is illustrated using a drop advection problem over the grids defined in Table \ref{tab:grid_transformation}. The initial condition is defined as a circle in the computational grid using Eq. \ref{eq:IC_advect}.
\begin{equation} \label{eq:IC_advect}
    \phi(\xi_1,\xi_2) = \frac{1}{2}\left(1-\tanh{\left(R - \sqrt{\frac{(\xi_1-\xi_1^c)^2+(\xi_2-\xi_2^c)^2}{2\epsilon_{\xi}}}\right)}\right)
\end{equation}
The radius of the circle, $R$, is defined as 0.25\% of the computational domain width and $\xi_1^c$ and $\xi_2^c$ are set as the centers of the computational domain. The time-step is set using an advection CFL of 0.1, and the phase field RHS parameters are chosen to be $\epsilon_{\xi} = 0.75 \Delta_{\xi}$ and $\Gamma_{\xi} = \max_i|U_i|$. The drop is advected for one period with a constant velocity of $u_1=5$ in physical space (final time of $t=0.2$). 
\begin{figure}[hbt!]
\begin{center}
\includegraphics[width=\textwidth]{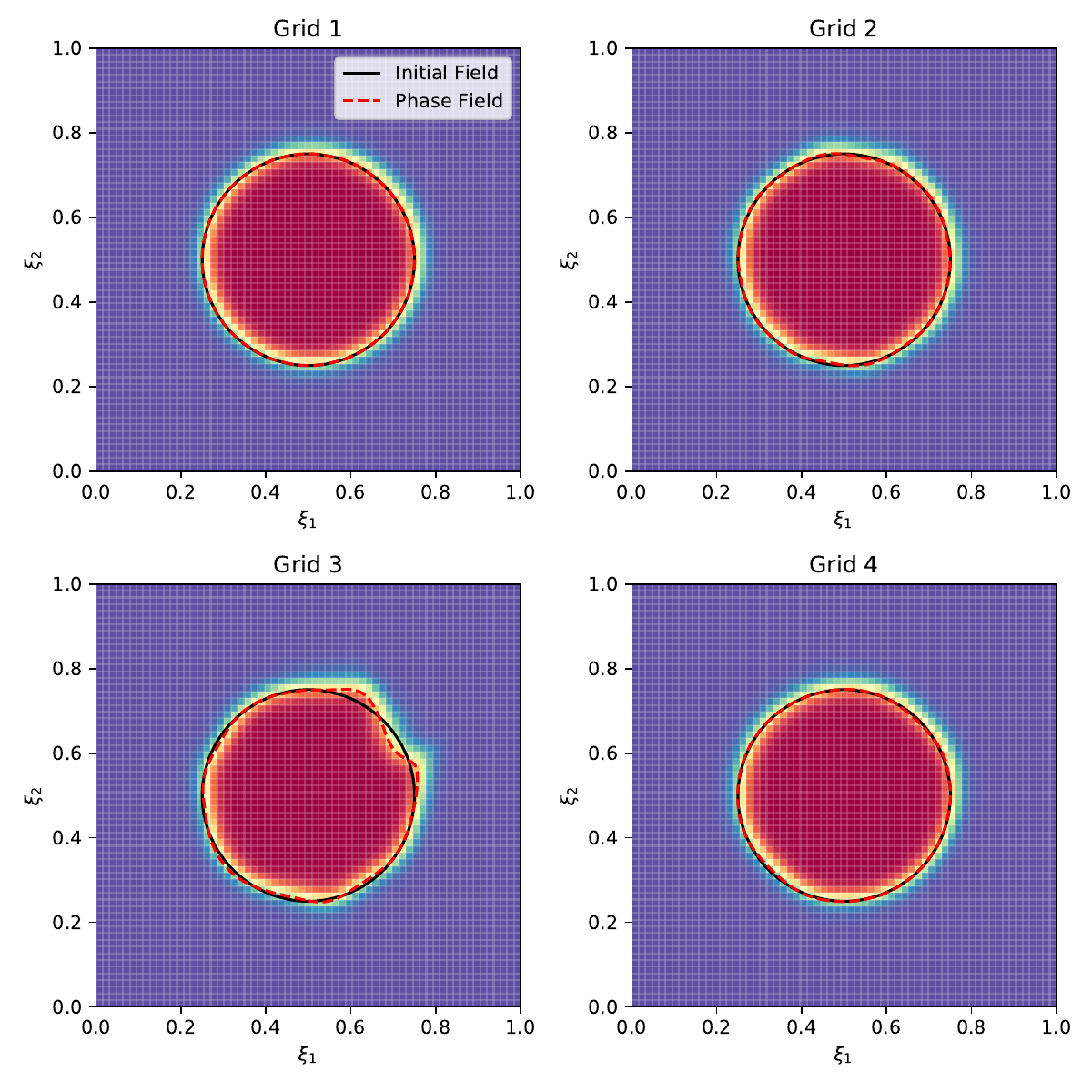}
\caption{Visual of a drop after one advection period for four different grids in computational space for the resolution $(N_x, N_y) = (64, 64)$. 
\label{fig:DropAdvection_computational}}
\end{center}
\end{figure}

\begin{figure}[hbt!]
\begin{center}
\includegraphics[width=\textwidth]{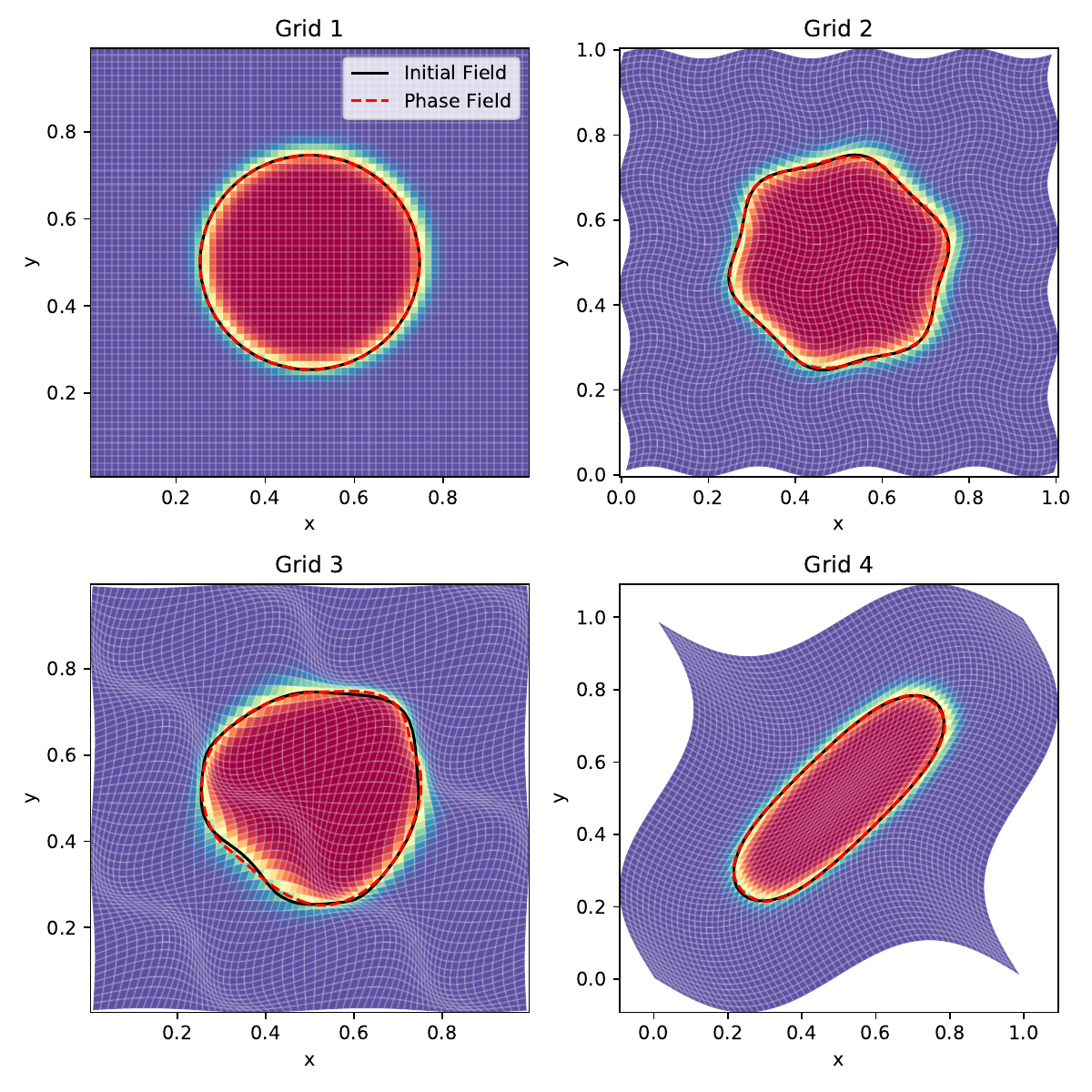}
\caption{Visual of a drop after one advection period for four different grids in physical space for the resolution $(N_x, N_y) = (64, 64)$.\label{fig:DropAdvection}}
\end{center}
\end{figure}

Fig. \ref{fig:DropAdvection_computational} shows the results in computational space and Fig. \ref{fig:DropAdvection}  shows the results in physical space. In computational space the drops are circular with a uniform interface thickness. The non-linear projection transforms the circular drops in computational space to non-circular shapes in physical space as shown in Fig. \ref{fig:DropAdvection}. The proposed projection also transforms the uniform interface in computational space to a grid-adapting interface thickness in physical space. The grid-adapting interface profile can be seen in the lower panels of Fig. \ref{fig:DropAdvection}. In particular, the lower left panel (Grid 3) visually illustrates the grid-adapting interface thickness near areas of varying grid resolution. 

\begin{figure}[hbt!]
\begin{center}
\includegraphics[width=\textwidth]{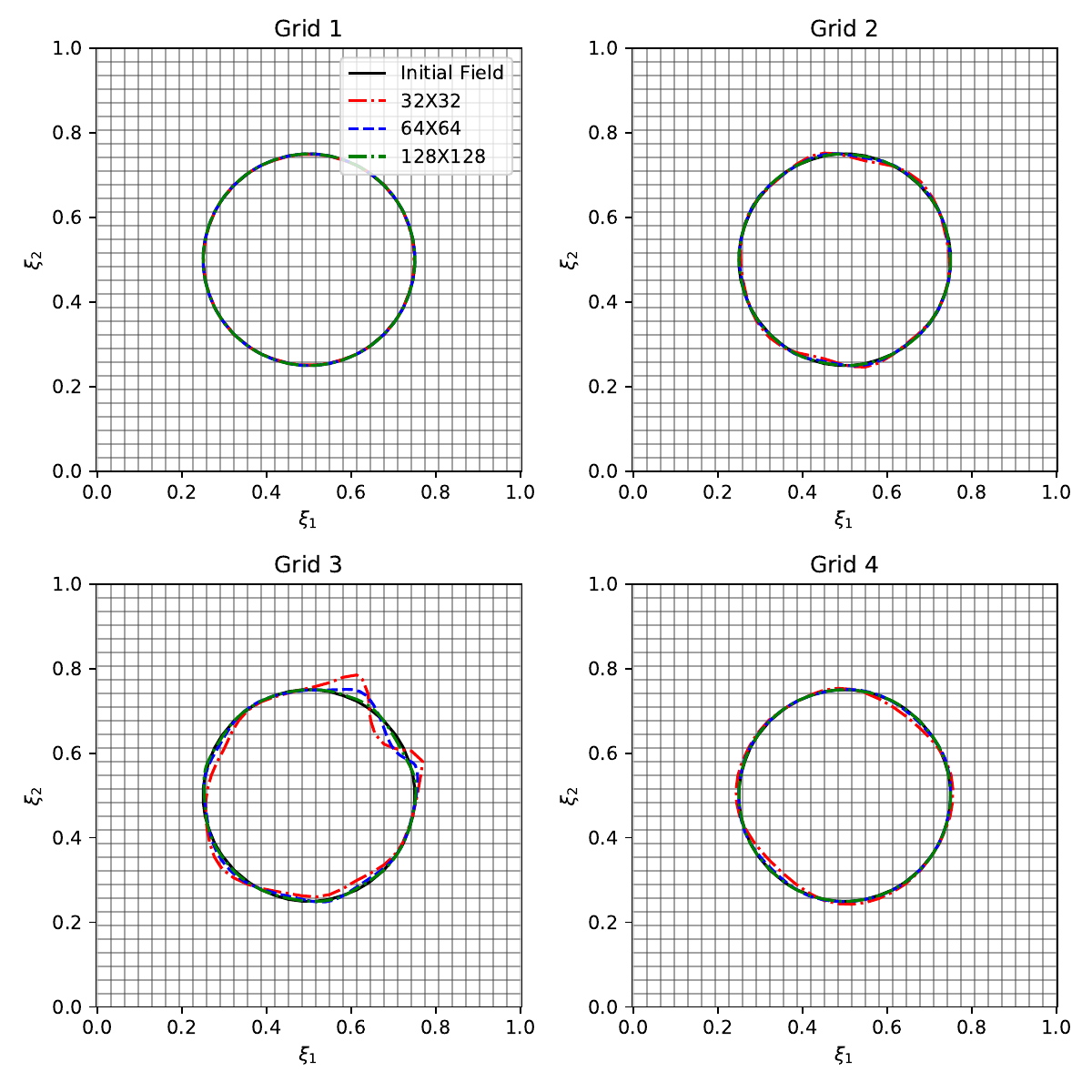}
\caption{Visual of the $\phi = 0.5$ contour of a drop after one advection period for four different grids in computational space. Convergence of the drop is plotted for three varying resolutions. All contours are overlayed over a sample coarse grid with resolution $(N_x, N_y) = (32, 32)$ to show the grid type. \label{fig:DropAdvection_comp_conv}}
\end{center}
\end{figure}

\begin{figure}[hbt!]
\begin{center}
\includegraphics[width=\textwidth]{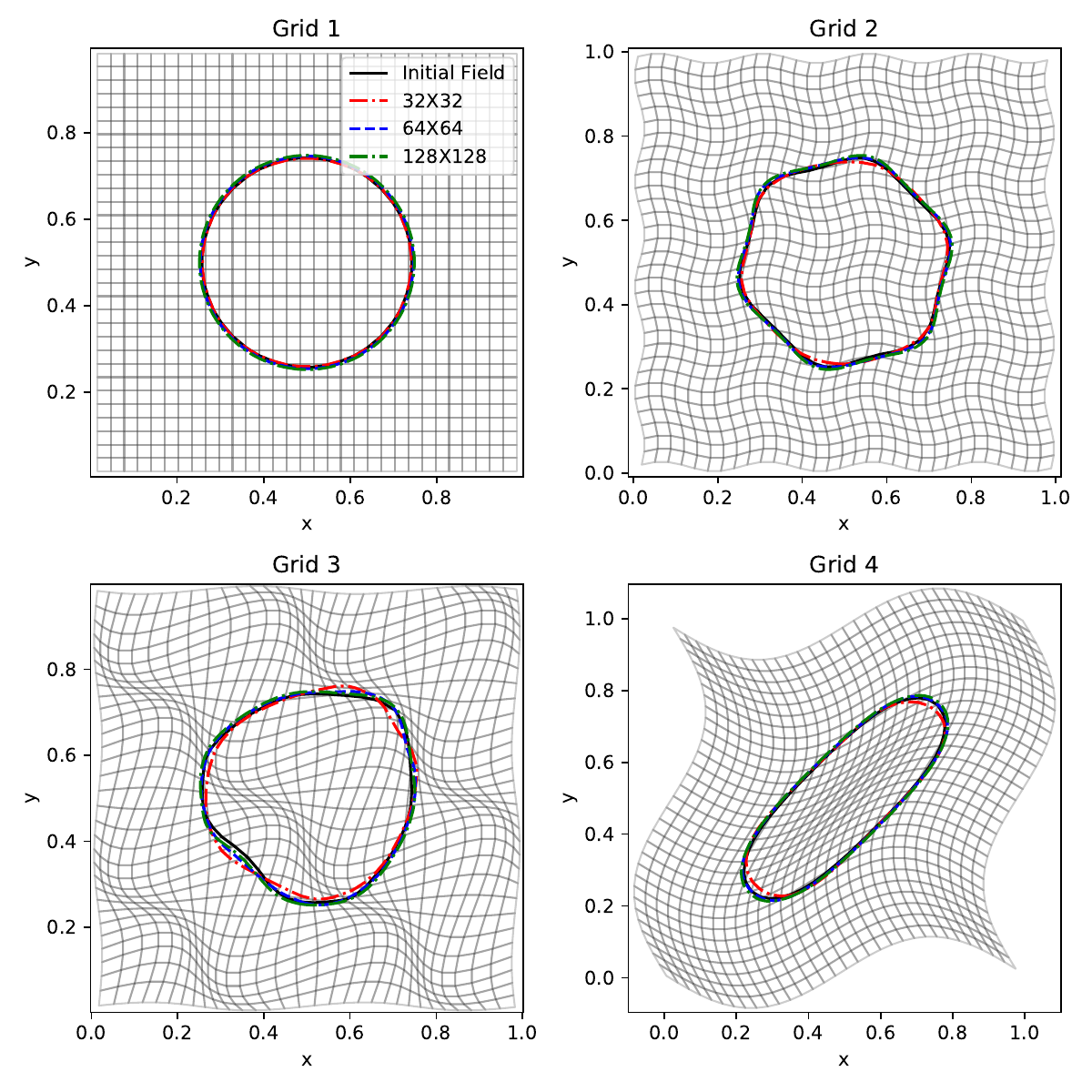}
\caption{Visual convergence of the $\phi = 0.5$ contour of a drop after one advection period for four different grids in physical space. Convergence of the drop is plotted using three varying resolutions. All contours are overlayed over a sample coarse grid with resolution $(N_x, N_y) = (32, 32)$ to show the grid type.
\label{fig:DropAdvection_phys_conv}}
\end{center}
\end{figure}

The constant and uniform velocity in the physical domain maps to a spatially non-uniform contravariant velocity, $U_i$, which deforms the drop in computational space throughout the spatial advection until returning the drop to the initial condition after a full period. The magnitude of the drop deformation throughout time is proportional to the skewness of the grid and causes larger interfacial shape error after an advection period. The convergence of the shape error was studied using grids with three resolutions, $(N_x \times N_y) = (32 \times 32)$, $(64 \times 64)$, and $(128 \times 128)$. The grid related error reduces with increasing grid resolution and the visual convergence of the 0.5 volume fraction contour of the drops in computational and physical space can be seen in Fig. \ref{fig:DropAdvection_comp_conv} and \ref{fig:DropAdvection_phys_conv} respectively. In computational space, Fig. \ref{fig:DropAdvection_comp_conv} shows that drops on all grids converge to the expected circular shape. In physical space, Fig. \ref{fig:DropAdvection_phys_conv} shows that the grids with larger skewness result in larger shape error. For instance, the curvilinear grid with the largest skewness (Grid 3) results in the phase field with the most visual deformation in both computational and physical space compared with the other grids.

\begin{figure}[hbt!]
\begin{center}
\includegraphics[width=0.7\textwidth]{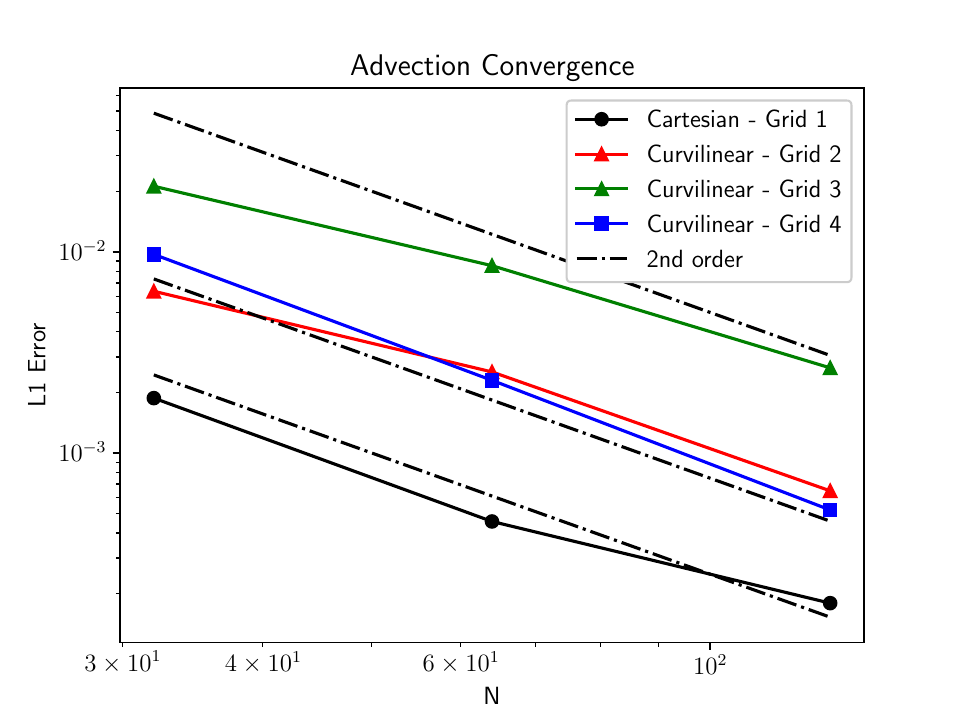}
\caption{L1 error convergence of a drop after one advection period for four different grids at varying resolutions. 
\label{fig:DropAdvection_Conv}}
\end{center}
\end{figure}

To quantitatively assess the error associated with the curvilinear grids, the L1 error is calculated by comparing the initial and final condition of the phase field equation using Eq. \ref{eq:L1_error}. 

\begin{equation} \label{eq:L1_error}
    \text{L1 Error} = \frac{\Sigma_{x,y}(|\phi(x,y)_\text{final} - \phi(x,y)_\text{initial}|)}{N_x \times N_y}
\end{equation}
The L1 convergence of the phase field on all grids is plotted in Fig. \ref{fig:DropAdvection_Conv} and converges near second order for all grids. The error magnitude increases with grid skewness which is consistent with the visual analysis of Fig. \ref{fig:DropAdvection_comp_conv} and \ref{fig:DropAdvection_phys_conv}. The error convergence for a general phase field problem contains two regimes: shape error and interfacial thickness error. In general, the shape error is larger at coarse resolutions whereas the interfacial thickness error is dominant at high resolutions. Convergence of shape error is dependent on the phase field method used with more accurate phase field approaches (e.g., CDI vs ACDI) converging at slightly different rates \cite{Jain2022ACDI}. On the other hand, the interface thickness error is directly dependent on the grid size and is expected to be a first order error for all phase field methods. As discussed, Figs. \ref{fig:DropAdvection_comp_conv} and \ref{fig:DropAdvection_phys_conv} show that the shape error is visible for the curvilinear grids and is expected to dominate the interfacial thickness error. On the other hand, the finest resolution for cartesian grid in Fig. \ref{fig:DropAdvection_Conv} begins to show a decrease in convergence rate, which is due to the magnitude of the  error decreasing below the magnitude of the interfacial thickness error. Further analysis of these two regimes is presented in \ref{appx:convergence}.

\subsection{Drop in Shear Flow on Curvilinear Grids} \label{sec:drop_shear}

\begin{figure}[hbt!]
\begin{center}
\includegraphics[width=\textwidth]{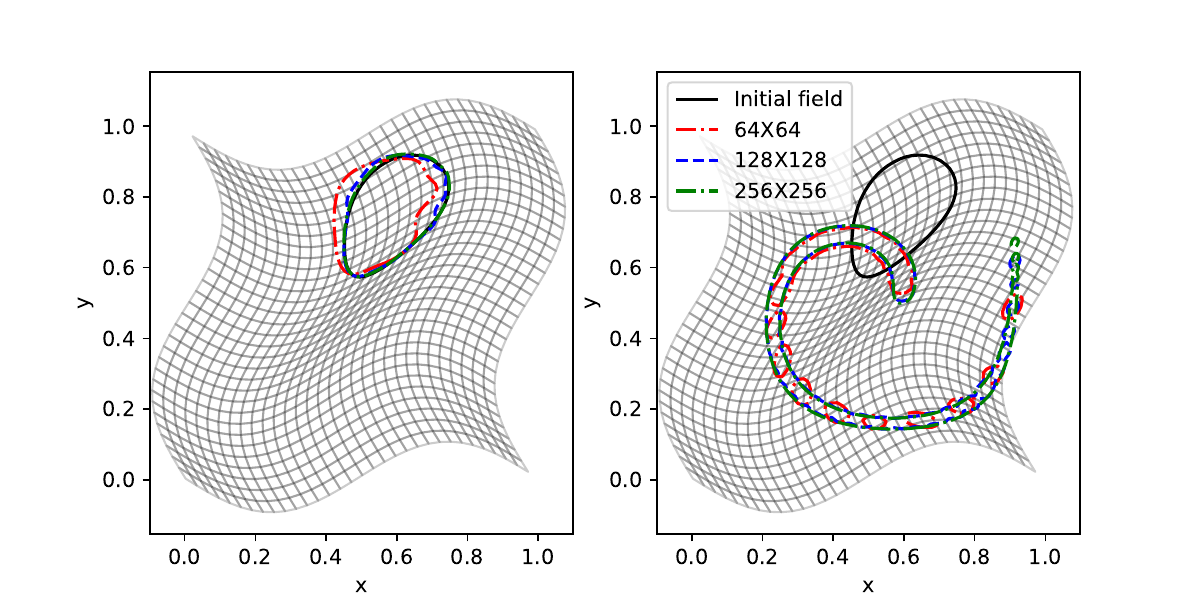}
\caption{Visual of the $\phi = 0.5$ contour of a drop deformation in shear flow on a skewed curvilinear grid (Grid 4, S = 1/10). The left shows convergence of the drop after one full period. The right shows the drop at one half period. All contours are overlayed over a sample coarse grid with resolution $(N_x, N_y) = (32, 32)$ to show the grid type.
\label{fig:DropInShear}}
\end{center}
\end{figure}

To assess the accuracy of the proposed curvilinear phase field projection consider a curvilinear version of the popular two-dimensional drop in shear flow test used to evaluate accuracy of interface reconstruction methods \cite{Rider1998,Chiu2011,Mirjalili2020}. Grid 4 is used for this test case and the system is initialized with a circular drop in the computational grid using Eq. \ref{eq:IC_advect} with $R=0.15\%$ of the computational domain centered at $\xi_1^c=0.5$, $\xi_2^c=0.75$. Similar to the advection case, the phase field equation is decoupled from the hydrodynamics equations and can be solved using an imposed velocity field defined in physical space using Eq. \ref{eq:Drop_in_shear},
\begin{equation}
    \begin{aligned} \label{eq:Drop_in_shear}
        u &= -\sin^2(\pi x)\sin (2\pi y)\cos \left(\frac{\pi t}{T}\right), \quad 
        v = \sin(2\pi x)\sin^2 (\pi y)\cos \left(\frac{\pi t}{T}\right) 
    \end{aligned}
\end{equation}
where $t$ is the simulation time, and $T=4$ is the time period of the flow. At $t=2$ the drop is at a maximally deformed state and the velocity field reverses to return the drop to its original shape at $t=T=4$. The time-step is set using an advection CFL of 0.1. The values of $\epsilon_{\xi} = 0.7$ and $\Gamma_{\xi} = \max_i|U_i|$ are used. The left panel of Fig. \ref{fig:DropInShear} shows the phase field contour at $t=4$, and the right panel shows the phase field at $t=2$. The drop profile at $t=4$ visually converges to the initial field with increasing grid resolution. 

\begin{figure}[hbt!]
\begin{center}
\includegraphics[width=0.7\textwidth]{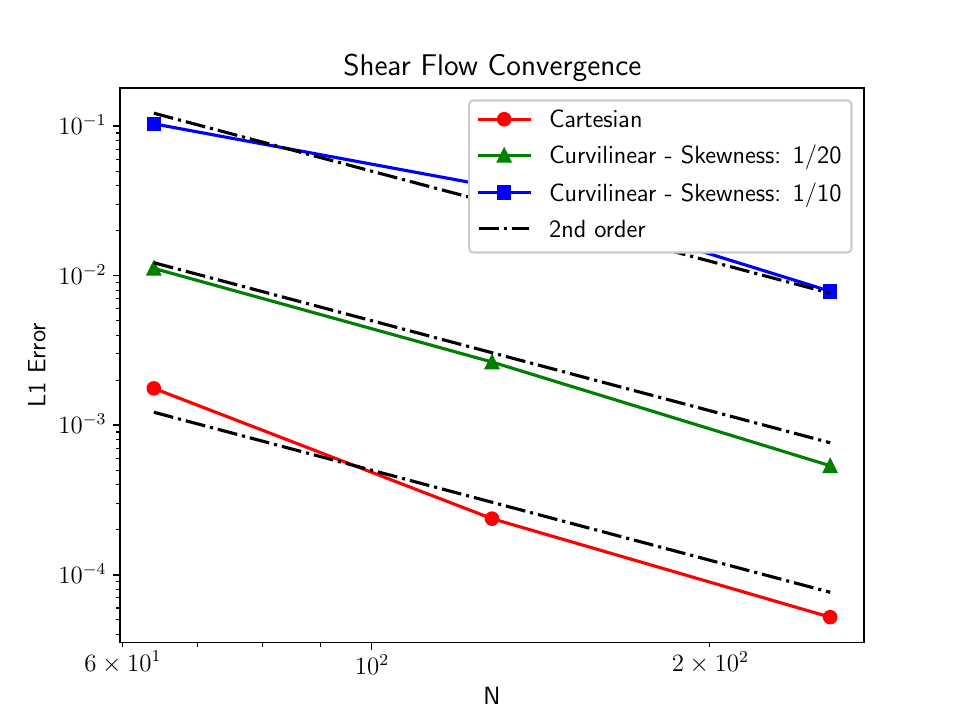}
\caption{L1 error convergence of drop in shear flow on curvilinear grids of varying skewness. \label{fig:Conv_DropInShear}}
\end{center}
\end{figure}

Using the initial field as the exact solution, an L1 error can be computed using Eq. \ref{eq:L1_error}. Three grids with varying skewness ($S = 0.0$, $1/20$, $1/10$) are discretized with resolutions, $(N_x \times N_y) = (64 \times 64)$, $(128 \times 128)$, and $(256 \times 256)$ to study the convergence for different curvilinear domains. Fig. \ref{fig:Conv_DropInShear} shows that the error for all of the configurations converge at 2nd order accuracy as the error is dominated by shape error for all grids. As expected, increasing the skewness parameter increases the magnitude of the error but does not impact the convergence rate of the solution. 

\subsection{Rayleigh Taylor}\label{sec:Results_RTI}

In this section, the phase field is two-way coupled to the hydrodynamics and a Rayleigh-Taylor instability is presented to evaluate the proposed curvilinear transformation for a physical two-phase simulation with non-unity density and viscosity ratios. To show that the proposed curvilinear extension of phase field models is physically consistent, a numerical comparison of converged (or nearly converged) high-resolution simulations on different physical grids is undertaken. The convergence of the instability is studied on all grid types listed in Table \ref{tab:grid_transformation}. 

Unlike the earlier tests in Section \ref{sec:Results} the full coupled system described in Section \ref{sec:NumericalMethods} is solved. To close this set of equations an equation of state must be defined. For this work a homogeneous 4-equation model equation of state (stiffened gas) is used, which assumes thermal and mechanical equilibrium (equal pressure, temperature, and velocity) between both fluids within a computational cell \cite{Flatten2011,lund2012hierarchy,lemartelot2013steady,le2014towards,saurel2016general}.  The complete thermodynamic state (including the volume fraction) can be found using the equation of state and the conserved variables without adding an additional transport equation for the volume fraction. 
The mixing rules are summarized below,
\begin{align}
    &T_1=T_2=T \\
    &p_1 = p_2 = p \\
    &e = Y_1e_1 + Y_2e_2 \\
    &\rho = \rho Y_1 + \rho Y_2 \\
\end{align}
where T is temperature, $p$ is pressure, $e_l$ is phasic internal energy, and $\rho_l$ is phasic density. Additionally, the following constraints are enforced,
\begin{align}
    &\phi_1 + \phi_2 = 1 \\
    &Y_1 + Y_2 = 1\\
    &\rho Y_l = \rho_l \phi_l
\end{align}
where $\phi_l$ is the phasic volume fraction. The stiffened gas equation of state for each phase, $l$, are,
\begin{align}
    &p_l(\rho_l,T_l) = \rho_l(\gamma_l-1)C_{v,l}T_l - p_{\infty,l} \\
    &e_l(p_l,T_l) = \frac{p_l+\gamma_lp_{\infty,l}}{p_l + p_{\infty,l}}C_{v,l}T_l \\
    &h_l(T_l) = \gamma_lC_{v,l}T_l
\\
    &c_l^2(P_l,\rho_l) = \frac{\gamma_l(p_l + p_{\infty,l})}{\rho_l}
\end{align}
where $C_v,l$ is the phasic isovolume specific heat capacity, $\gamma_l$ is the ratio of phasic specific heats, and $p_{\infty,l}$ is the stiffened pressure representing the attractive effects in the liquid phase. Combining the phase EOS with the mixing rules closes the system with,
\begin{equation}
    p = \frac{-a_1+\sqrt{a_1^2-4a_0a_2}}{2a_2}
\end{equation}
where
\begin{align}
    a_0 &= Y_1C_{v,1}+Y_2C_{v,2} \\
        a_1 &= Y_1C_{v,1}(p_{\infty,2} + \gamma_1p_{\infty,1}-(\gamma_1-1)\rho e) + Y_2C_{v,2}(p_{\infty,1} + \gamma_2p_{\infty,2}-(\gamma_2-1)\rho e) \\
        a_2 &= -\rho e((\gamma_1-1)Y_1C_{v,1}p_{\infty,2} + (\gamma_2-1)Y_2C_{v,2}p_{\infty,1}) + p_{\infty,1}p_{\infty,2}(\gamma_1Y_1C_{v,1} + \gamma_2Y_2C_{v,2})
\end{align}
For this work all material parameters, ($\gamma_l$, $p_{\infty,l}$, and $C_{v,l}$) are assumed constant.  

The Rayleigh-Taylor instability with a density ratio $\rho_1/\rho_2 = 5$, dynamic viscosity ratio $\mu_1/\mu_2 = 10$, and surface tension is ignored. The acceleration due to gravity force is set to 9.81. The RTI is simulated on a variety of grids to study the impact of the curvilinear phase field transformation. The physical domain defined in Table \ref{tab:grid_transformation} and \ref{tab:grid_params} is initialized with initial fluid profile given by Eq. \ref{eq:RT_IC}.
\begin{equation} \label{eq:RT_IC}
    \phi_1 = \frac{1}{2}\left(1+\tanh\left(\frac{y-h-0.1\cos\left(2\pi x\right)}{2 (0.75 w/N)}\right)\right)
\end{equation}
where $h = 2$ and $w = 3$ and $N$ is the number of grid points $(N_x = N_y = N)$. The initial hydrodynamic and thermodynamic state is initialized with $\Vec{u}=0$, $T=1$ and $p=1000$. The stiffened gas parameters for phase 1 are taken as, $p_{\infty,1} = 800$, $\gamma_1 = 3$, $C_{v,1} = 180$, and $\mu_1=0.1$. For phase 2, $p_{\infty,2} = 0$, $\gamma_2 = 1.4$, $C_{v,2} = 2500$, and $\mu_2=0.01$. The phase field parameters are defined on the computational grid with $\epsilon_{\xi} = 0.75\Delta_{\xi}$ and $\Gamma_{\xi} = \max_i|U_i|$. Periodic boundary conditions are used in the x-direction and adiabatic walls are used in the y-direction. The simulations are run until $t=1.0$ with an advection CFL = 0.5. 

\begin{figure}[hbt!]
\begin{center}
\includegraphics[width=\textwidth]{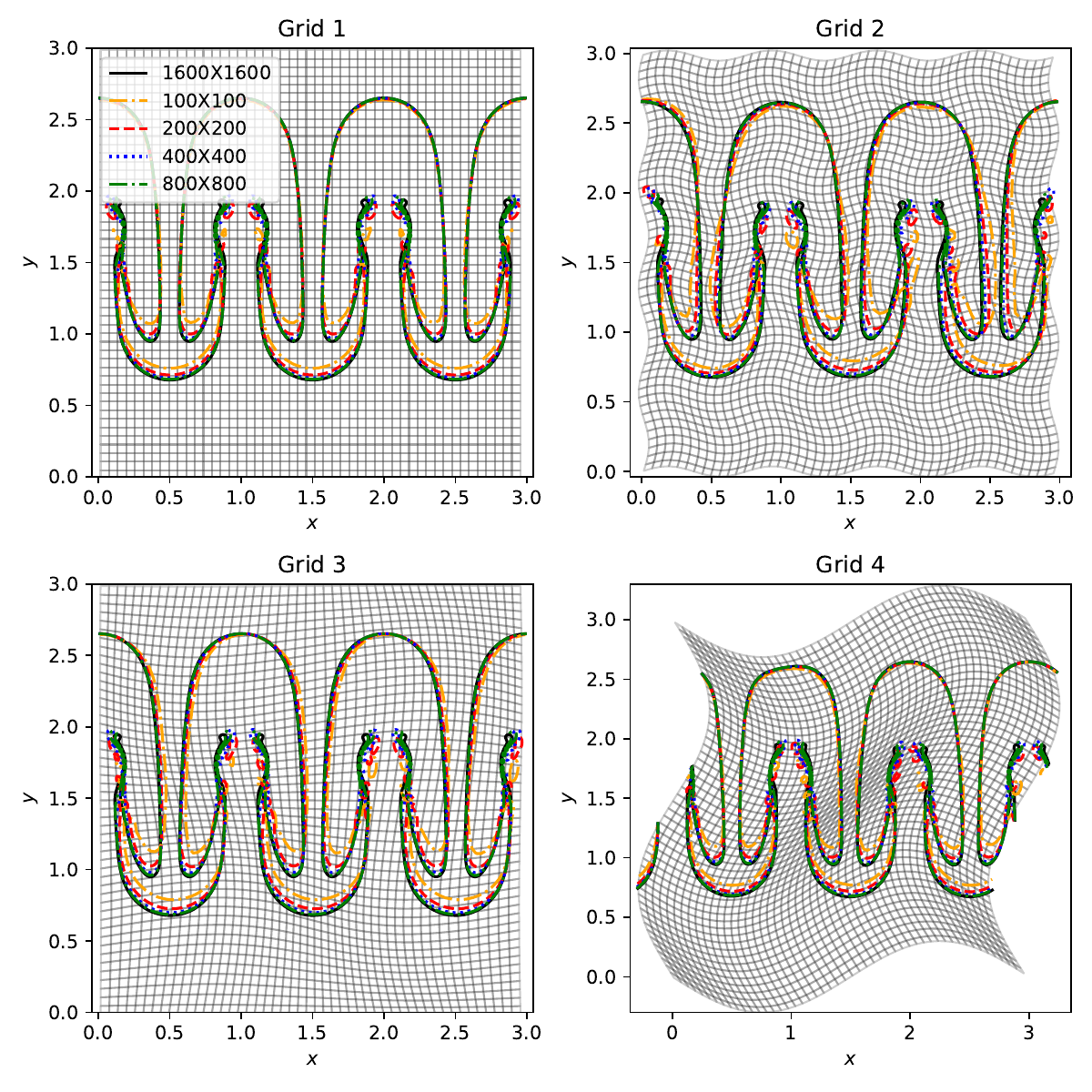}
\caption{Raleigh Taylor simulation on four different grids.
Visual convergence of the $\phi = 0.5$ contour of a Rayleigh-Taylor instability at $t=1.0$ for four different grids in physical space. Convergence of the contour is plotted using five varying resolutions. All contours are overlayed over a sample coarse grid with resolution $(N_x, N_y) = (50, 50)$ to show the grid type.
\label{fig:RT_instability_physcical}}
\end{center}
\end{figure}

The $\phi_1 = 0.5$ contour of the phase field solutions of the Rayleigh-Taylor simulations can be seen on grids in physical space in Fig. \ref{fig:RT_instability_physcical}. 
At low resolutions, there is numerical breakup caused by the phase field terms enforcing finite and resolvable interface thickness. As resolution is increased the interfacial features become more resolved and less artificial breakup is observed. For medium level resolutions this results in small drops along the path of the instability ligament. With sufficient resolution these small interfacial features are resolvable and no breakup occurs (as expected for a two-dimensional problem without surface tension). 

\begin{figure}[hbt!]
\begin{center}
\includegraphics[width=\textwidth]{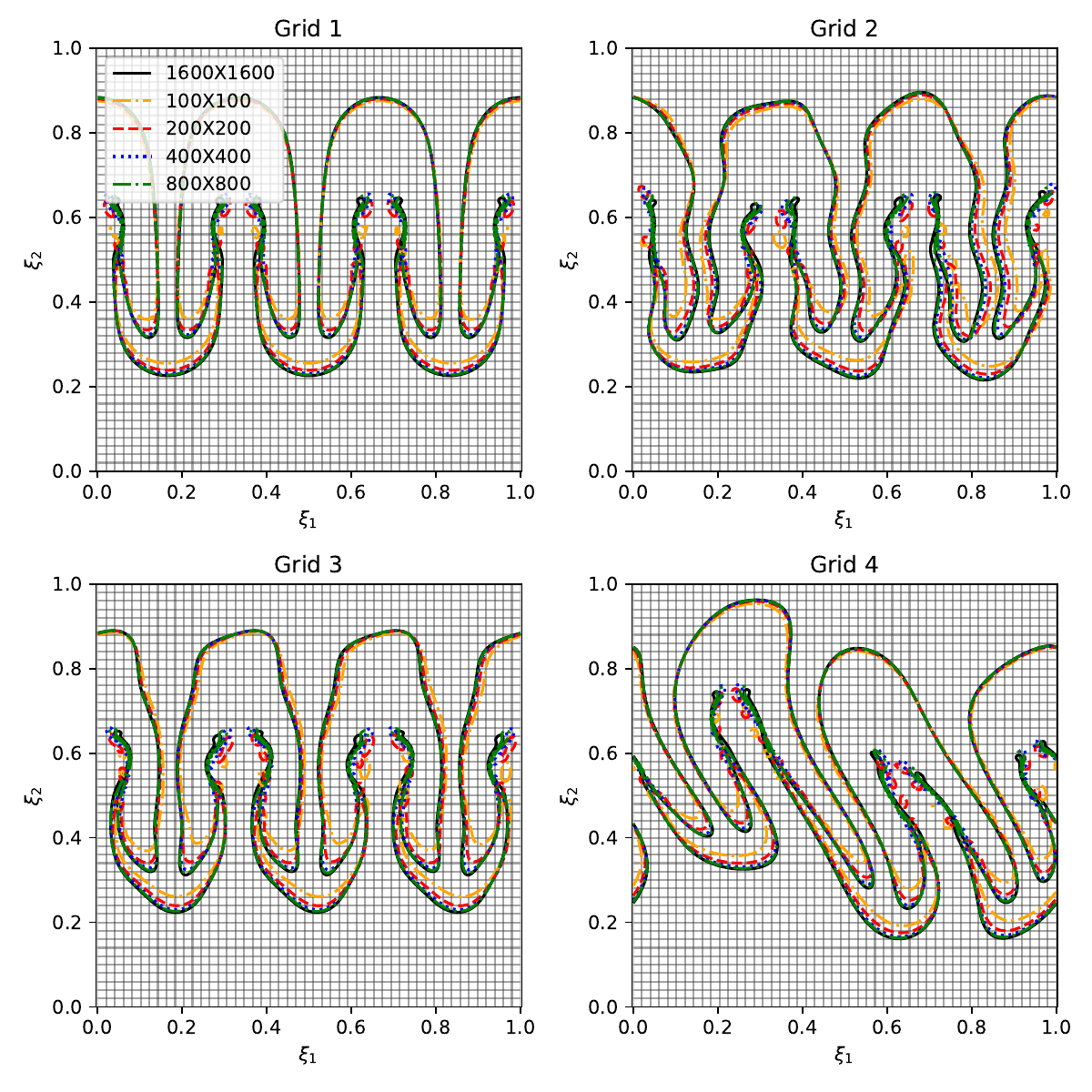}
\caption{Visual convergence of the $\phi = 0.5$ contour of a Rayleigh-Taylor instability at $t=1.0$ for four different grids in computational space. Convergence of the contour is plotted using five varying resolutions. All contours are overlayed over a sample coarse grid with resolution $(N_x, N_y) = (50, 50)$ to show the grid type.
\label{fig:RT_instability_computational}}
\end{center}
\end{figure}

\begin{figure}[hbt!]
\begin{center}
\includegraphics[width=0.2\textwidth]{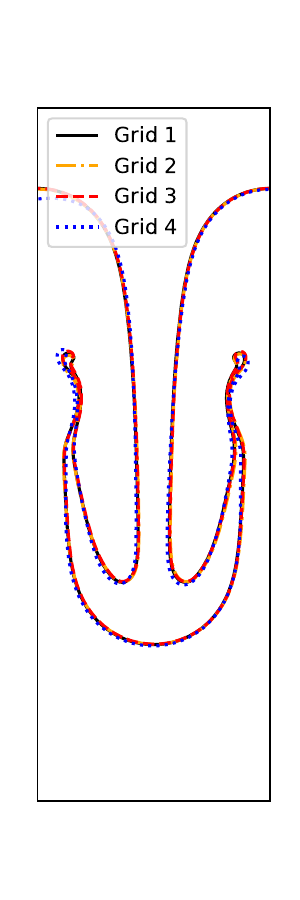}
\caption{Visual convergence of the $\phi = 0.5$ contour of the center Rayleigh-Taylor instability at $t=1.0$ between four different curvilinear grids at the resolution $(N_x, N_y) = (1600, 1600)$. 
\label{fig:RT_grid_comparison}}
\end{center}
\end{figure}

This is the first numerical experiment in this work where the the phase field regularization is two-way coupled with the hydrodynamics. In past works on isotropic cartesian grids, the phase field operator acted directly on the grid which transports the hydrodynamic variables. For the proposed curvilinear transformation, the regularization of the phase interface is done in the computational domain which can have drastically different interface shapes than the physical. Fig. \ref{fig:RT_instability_computational} shows the same Rayleigh-Taylor solutions at $t=1.0$ in the computational grid where the solutions display noticeably different characteristics. To ensure that applying the phase field regularization in computational space does not negatively influence the hydrodynamics and impact computational consistency, Fig. \ref{fig:RT_grid_comparison} overlays the 0.5 volume fraction contour of the center instability from the highest resolution simulations completed in Fig. \ref{fig:RT_instability_physcical} $(N_x = 1600, N_y = 1600)$. Fig. \ref{fig:RT_grid_comparison} confirms that all grids visually converge to the same solution. The slight differences shown for the result from Grid 4 can be attributed to the earlier interactions of the top wall with the instability (the leftmost bubble gets visibly closer to the top wall compared to all other grids). The proposed curvilinear projection results in a grid independent converged solution providing further evidence that our proposed approach for curvilinear grid treatment is physically consistent.  

\section{Conclusion} \label{sec:Conclusion}

In this work, we proposed a framework for transforming popular phase field methods to generalized curvilinear coordinates. Applying the phase field regularization in computational space produces a constant interface thickness in the computation domain which projects to a grid-adapting interface thickness in the physical grid. Additionally, proof of boundedness of volume fraction for the CDI model using second-order centered schemes has been extended to the curvilinear domain using the curvilinear form of CDI (similar proofs for ACDI can be carried out as well). The proposed form of the ACDI curvilinear phase field model was tested with three numerical experiments on multiple curvilinear grids. All experiments showed the proposed transformation creates stable and accurate phase interfaces. For all cases, including highly skewed grids, the convergence rate is near 2nd order and identical for both cartesian and curvilinear grids. Lastly, the proposed approach was fully coupled to the compressible Navier-Stokes equations and a Rayleigh-Taylor instability was solved using four grids of varying skewness. The phase field visibly converged with increasing resolution for each grid. Additionally, solutions on all grids converged to near identical solutions in physical space, showing the solution of the proposed curvilinear transformation is consistent. 

\section*{Acknowledgments}
This work is based upon work supported by the Department of Energy, National Nuclear Security Administration under Award Number DE-NA0003968 within the PSAAP III (INSIEME) Program at Stanford University. We are also grateful for the insightful discussions with Dr. Makrand Khanwale, Deniz Bezgin, and Reed Brown.

\appendix
\section{Curvilinear Transformations for Common Phase Field Approaches} \label{appx:transformations}

The same approach described in detail in Section \ref{sec:Proposed_curvilinear_form}, can be applied to other popular phase field methods. Table \ref{tab:Form_Summary} shows the forms for CDI, ACDI, conservative Allen Cahn, and conservative Cahn Hilliard with variable mobility. 

\begin{table}[hbt!]
    \centering
    \begin{tabular}{l l l}
    \hline
    \\
    Name & Cartesian: $g(\phi,x_i)$  & Curvilinear: $G(\phi,\xi_i)$ \\
    \\
    \hline
    \\
    CDI \cite{Chiu2011,Mirjalili2020,Mirjalili2021,Jain2020} & $\frac{\partial}{\partial x_i}\left[\Gamma\left(\epsilon\frac{\partial \phi}{\partial x_i} - \phi(1-\phi)\frac{\partial \phi/\partial x_i}{|\Vec{\nabla}\phi|}\right)\right]$ &
    $\frac{\partial }{\partial \xi_i}\left[\Gamma_{\xi}\left(\epsilon_{\xi}\frac{\partial \phi}{\partial \xi_i} - \phi(1-\phi)\frac{\partial \phi/\partial \xi_i}{|\Vec{\nabla}\phi|}\right)/J\right]$   \\
    \\
    \hline
    \\
    ACDI \cite{Jain2022ACDI} & $\frac{\partial}{\partial x_i}\left[\Gamma\left(\epsilon\frac{\partial \phi}{\partial x_i} - \frac{1}{4}\left[1-\tanh^2{\left(\frac{\psi}{2\epsilon}\right)}\right]\frac{\partial \psi/\partial x_i}{|\Vec{\nabla}\psi|}\right)\right]$ &
    $\frac{\partial }{\partial \xi_i}\left[\Gamma_{\xi}\left(\epsilon_{\xi}\frac{\partial \phi}{\partial \xi_i} - \frac{1}{4}\left[1-\tanh^2{\left(\frac{\psi_{\xi}}{2\epsilon_{\xi}}\right)}\right]\frac{\partial \psi_{\xi}/\partial \xi_i}{|\Vec{\nabla}\psi_{\xi}|}\right)/J\right]$  \\
    \\
    \hline
    \\
    AC \cite{allen1979,lee2016comparisonAC_CH} & $\frac{\partial}{\partial x_j}\left[ \epsilon^2 \frac{\partial \phi}{\partial x_j}\right] - F'(\phi(x_i,t)) + \alpha(t)$ & $\frac{\partial }{\partial \xi_i}\left[\epsilon_{\xi}^2\frac{\partial \phi}{\partial \xi_i}/J\right] - F'(\phi(x_i,t)) + \alpha(t)$  \\
    \\
    \hline
    \\
    CH \cite{cahn1958,lee2016comparisonAC_CH} & $\frac{\partial}{\partial x_i}\left[M(\phi)\frac{\partial }{\partial x_i}\left(F' - \epsilon^2\frac{\partial^2 \phi}{\partial x^2}\right)\right]$ & $\frac{\partial }{\partial \xi_i}\left[M(\phi)\frac{\partial }{\partial \xi_i}\left(F'(\phi(\xi,t)) - \epsilon_{\xi} \frac{\partial^2 \phi}{\partial \xi_j\xi_j}\right)/J\right]$  \\
    \\
    \hline
    \end{tabular}
    \caption{Summary of transformations from cartesian to curvilinear grids for popular phase field methods}
    \label{tab:Form_Summary}
\end{table}

\section{A note on boundedness} \label{sec:boundedness}
Following past works from \cite{Mirjalili2020,Jain2020}, we can show that by using the current model, the phase field variable remains bounded for certain choices of $\epsilon_{\xi}$ and $\Gamma_{\xi}$ on curvilinear grids using second order schemes. We seek to show that provided,
\begin{align} \label{eq:CDI_epsilon}
    \frac{\epsilon_{\xi}}{\Delta \xi} \geq \frac{|U|_{max}/\Gamma_{\xi} + 1}{2},
\end{align}
and
\begin{align} \label{eq:CDI_deltaT}
    \Delta t \leq \frac{\Delta \xi^2}{2\Gamma_{\xi} \epsilon_{\xi}},
\end{align}
the phase field will stay bounded between $0 \leq \phi \leq 1$. As explained in \cite{Mirjalili2020}, the most restrictive case to prove boundedness of the phase field variable is explicit Euler in one spatial dimension. Extensions to higher dimensional flows is straightforward and examples are shown in past work \cite{Mirjalili2020}. The discretization of Eq. \ref{eq:PF_curvilinear} where $G$ is defined using Eq. \ref{eq:CDI_comp_curv} on a one-dimensional grid in computational space with explicit Euler time advancement is given by,

\begin{equation} \label{eq:discrete_CDI}
    \begin{split} 
    \phi_i^{k+1} &= \phi_i^{k} + \Delta t\Bigg[-\frac{U_{i+1}^k\phi_{i+1}^k/J_{i+1} - U_{i-1}^k\phi_{i-1}^k/J_{i-1}}{2\Delta\xi} + \\  &  \Gamma_{\xi} \epsilon_{\xi} \Bigg\{\frac{\phi_{i-1}^k/J_{i-1} - 2\phi_i^k/J_i + \phi_{i+1}^k/J_{i+1}}{\Delta \xi^2} - \\
    &\Gamma_{\xi}\Bigg(\frac{(1-\phi_{i+1}^k)n_{i+1}^k\phi_{i+1}^k/J_{i+1} - (1-\phi_{i-1}^k)n_{i-1}^k\phi_{i-1}^k/J_{i-1}}{2\Delta\xi}\Bigg)\Bigg\}\Bigg]
    \end{split}
\end{equation}
where $k$ represents the time-step index and $i$ is the computational grid index. Equation \ref{eq:discrete_CDI} can be arranged as,

\begin{align}
    \phi_i^{k+1} = C_{i-1}\phi_{i-1}/J_{i-1} + C_i\phi_i/J_i + C_{i+1}\phi_{i+1}/J_{i+1}
\end{align}
where,
\begin{align}
    &C_{i-1} = \frac{\Delta tU_{i-1}^k}{2\Delta \xi} + \frac{\Delta t\Gamma_{\xi} \epsilon_{\xi}}{\Delta \xi^2} + \frac{\Delta t \Gamma_{\xi}}{2 \Delta \xi}(1 - \phi_{i-1}^k)n_{i-1}^k, \\
    &C_{i+1} = -\frac{\Delta tU_{i+1}^k}{2\Delta \xi} + \frac{\Delta t\Gamma_{\xi} \epsilon_{\xi}}{\Delta \xi^2} - \frac{\Delta t \Gamma_{\xi}}{2 \Delta \xi}(1 - \phi_{i+1}^k)n_{i+1}^k, \\
\end{align}
and 
\begin{align}
    &C_{i} = 1 - \frac{2 \Delta t \Gamma_{\xi} \epsilon_{\xi}}{\Delta \xi^2}
\end{align}

Assuming all $\phi$ values are positive at time step $k$ ($\phi_i^k\ge0$ for all $i$), we seek to prove that if Eqs. \ref{eq:CDI_epsilon} and Eq. \ref{eq:CDI_deltaT} are satisfied, then $\phi_i^{k+1}\ge0$ for all $i$ indices too. For an admissible grid (non-overlapping cells), $J_i > 0$, so we need to show that all $C_i$ values are positive given the constraints given by Eqs. \ref{eq:CDI_epsilon} and Eq. \ref{eq:CDI_deltaT}. We can see immediately from the time-step restriction that $C_i \geq 0$. Additionally, we can note that by definition, the interface normal, $n_{i-1,i+1}^k \geq -1$. With this fact, we can see that $C_{i-1} \geq \Delta t U_{i-1}^k/(2\Delta \xi_1) + \Delta t \Gamma_{\xi} \epsilon_{\xi}/\Delta\xi^2 - \Delta t \Gamma_{\xi}/(2\Delta \xi) \geq -\Delta t(|U|_{max}^k + \Gamma_{\xi})/(2\Delta \xi) + \Delta t\Gamma_{\xi} \epsilon_{\xi}/\Delta \xi_1^2$. Now, using Eq. \ref{eq:CDI_epsilon} we can see that $C_{i-1} \geq 0$. A similar analysis can be done for $C_{i+1}$. All together, we have shown that if Eq. \ref{eq:CDI_epsilon} and Eq. \ref{eq:CDI_deltaT} are satisfied, then $\phi$ remains positive throughout time with the proposed  curvilinear form of CDI. To prove that $\phi\le1$, the same analysis carried out above can be applied to $\tilde{\phi}=1-\phi$, whereby one can show that $\tilde{\phi}$ remains positive. Altogether, we have outlined the proof for the boundedness of the proposed model in a one-dimensional setting with explicit Euler time integration. As mentioned, the one dimensional case with explicit Euler is the most restrictive case for proving boundedness and extensions to higher dimensions can be found in past work \cite{Mirjalili2020}. This proof follows the past works assuming second-order central discretizations of all terms. In this work, as well as past works, higher order schemes have been used for the advection term. Though no formal proof currently exists the boundedness properties for high-order schemes, the constraints derived assuming a second-order central discretization have held true for high-order schemes in practice (including the tests shown in this work).

\section{Convergence Regimes for Phase Field approaches} \label{appx:convergence}

\begin{figure}[hbt!]
\begin{center}
\includegraphics[width=0.7\textwidth]{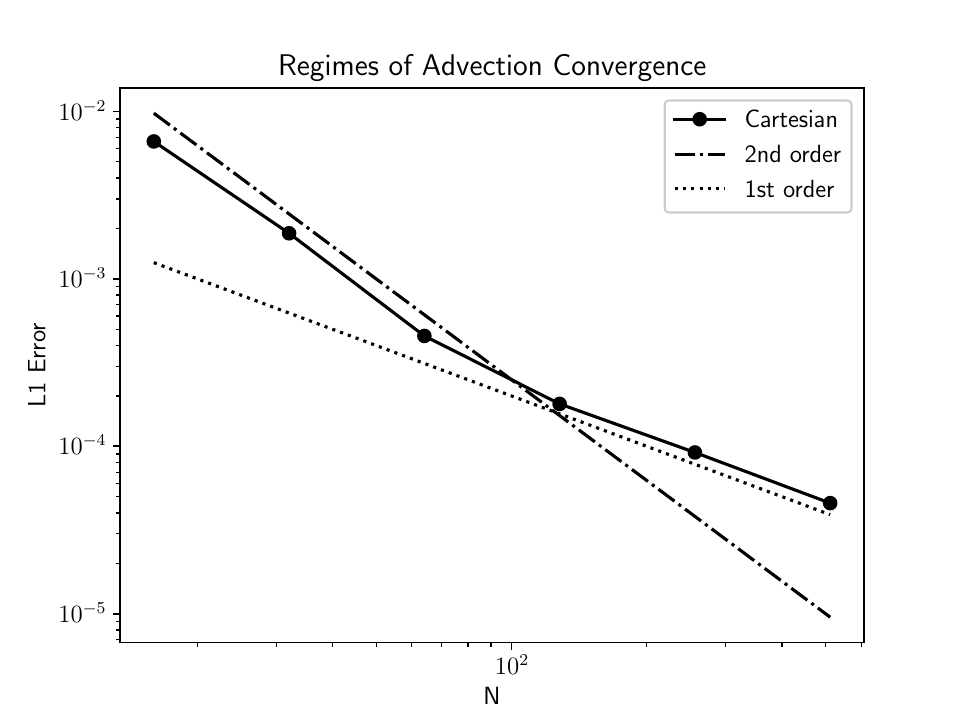}
\caption{Convergence of a drop after one advection period on a cartesian grid. Shape error (2nd order) dominates at low resolution and interfacial error (1st order) dominates at high resolution. \label{fig:Convergence_regimes}}
\end{center}
\end{figure}

The error convergence for a general phase field problem contains two regimes: a regime dominated by shape error and a regime dominated by interfacial thickness error. For coarse resolutions, advection of an interface will result in deformation which contributes the majority of the phase field shape error. In this regime increasing grid resolution results in more accurate interfacial shapes which provides error convergence. Figs. \ref{fig:DropAdvection_Conv} and \ref{fig:Conv_DropInShear} are examples of convergence plots within the shape error regime. For both of these cases the L1 error decreases with near 2nd order convergence for all grids. To clearly illustrate the existence of these regimes, a detailed convergence study was completed using a drop advection problem on a cartesian grid. Lower values of N in Fig. \ref{fig:Convergence_regimes} show 2nd order convergence associated with interface deformation. As shown in Fig. \ref{fig:Convergence_regimes}, when the drop was resolved with around 50 points per diameter, the convergence rate began to decrease. At this level of resolution the shape error became much smaller than the interfacial thickness error associated with the finite thickness of a phase field method. The interfacial thickness is determined by the phase field parameter, $\epsilon$, which is proportional to the grid resolution by definition. As expected, the interfacial thickness error will result in a first-order converging error with increasing grid resolution. For this advection test on a uniform cartesian grid values of N leading to O(100) points per droplet diameter are dominated by interface thickness error and converge at 1st order. 

The amount of points required to switch from the shape error regime to the interfacial thickness regime is dependent on both the hydrodynamic flow field and the grid skewness. The advection problem in Section \ref{sec:Drop_Advection} shows that a highly skewed grid will induce more shape error compared to a uniform grid. Furthermore, the drop in shear test case from Section \ref{sec:drop_shear} illustrates that a hydrodynamic flow field which cause large interface straining will result in higher shape error. Given these observations, the error associated with the phase field for a complex hydrodynamics problem on a general grid is likely to be dominating by shape error.

\section{Transforming Operators to Generalized Curvilinear Coordinates} \label{appx:OperatorCurvilinearTransformation}

\subsection{Converting derivative of scalar in physical to curvilinear coordinates}

Converting the derivative of a scalar from one field to another is done using the chain rule,

\begin{equation}
    \frac{\partial \phi}{\partial x} = \frac{\partial \phi}{\partial \xi_i}\frac{\partial \xi_i}{\partial x} = \frac{\partial \phi}{\partial \xi_1}\frac{\partial \xi_1}{\partial x} + \frac{\partial \phi}{\partial \xi_2}\frac{\partial \xi_2}{\partial x} + \frac{\partial \phi}{\partial \xi_3}\frac{\partial \xi_3}{\partial x}. \label{eq:divg_scalar}
\end{equation}

\subsection{Converting divergence of vector in physical to curvilinear coordinates}

Converting a directionally dependent vector field from physical coordinates to computational coordinates requires projecting the vector in physical space to the computational space. This can be done using the inner product between the physical vector and the surface-normal area vector as,

\begin{equation}
    \frac{\partial \phi_i}{\partial x_i} = J\frac{\partial (\phi_j  S_{ji})}{\partial \xi_i} = J\left[ \frac{\partial}{\partial \xi_1}\left(\frac{1}{J}\left(\phi_j\frac{\partial \xi_1}{\partial x_j} \right)\right) + \frac{\partial}{\partial \xi_2}\left(\frac{1}{J}\left( \phi_j\frac{\partial \xi_2}{\partial  x_j} \right)\right) + \frac{\partial}{\partial \xi_3}\left(\frac{1}{J}\left( \phi_j\frac{\partial \xi_3}{\partial  x_j} \right)\right)\right]\label{eq:divg_vec}
\end{equation}
where 
\begin{align}
    S_{ji} = \frac{1}{J}\frac{\partial \xi_i}{\partial x_j} =\frac{1}{J}
\begin{bmatrix}
    \frac{\partial \xi_1}{\partial x} \quad \frac{\partial \xi_1}{\partial y} \quad \frac{\partial \xi_1}{\partial z} \\
    \frac{\partial \xi_2}{\partial x} \quad \frac{\partial \xi_2}{\partial y} \quad \frac{\partial \xi_2}{\partial z} \\
    \frac{\partial \xi_3}{\partial x} \quad \frac{\partial \xi_3}{\partial y} \quad \frac{\partial \xi_3}{\partial z} \\
\end{bmatrix}^T
\end{align}
and ${1}/{J} = \det\left( {\partial x_i}/{\partial \xi_j}\right)$.


\bibliographystyle{elsarticle-num}
\bibliography{elsarticle-template}

\end{document}